\documentclass[amssymb,aps,prl,preprint,groupedaddress,showpacs]{revtex4}

\usepackage{graphicx}

\begin{document}

\title{Comment on `Stretched polymers in a poor solvent'}
\author{A. L. Owczarek}
\email{aleks@ms.unimelb.edu.au}
\affiliation{Department of Mathematics and Statistics, The University of Melbourne, 3010, Australia}
\author{T. Prellberg}
\email{thomas.prellberg@tu-clausthal.de}
\affiliation{Institut f\"ur Theoretische Physik, Technische Universit\"at Clausthal, Arnold Sommerfeld Stra\ss e 6, D-38678 Clausthal-Zellerfeld, Germany}

\date{\today }

\begin{abstract} 
In a recent article on stretched polymers in a poor solvent by
Grassberger and Hsu \cite{grassberger2002a-a} questions were raised as
to the conclusions that can be drawn from currently proposed scaling
theory for a single polymer in various types of solution in two and
three dimensions. Here we summarise the crossover theory predicted for
low dimensions and clarify the scaling arguments that relate thermal
exponents for quantities on approaching the $\theta$-point from low
temperatures to those associated with the asymptotics in polymer
length at the $\theta$-point itself.
\end{abstract}

\maketitle

Recently some interesting work has been completed on stretched
polymers in a poor solvent by Grassberger and Hsu
\cite{grassberger2002a-a} (GH) and on collapsed polymers on a cylinder
by Hsu and Grassberger \cite{hsu2002a-a} (HG).  In the course of these
works various scaling conjectures were discussed; the question arose
as to whether they can be derived from theory currently in the
literature.

The basic framework of the polymer problem has been, and still is,
given by the seminal works of De Gennes and Duplantier
\cite{gennes1975a-a,stephen1975a-a,duplantier1982a-a} which describe
the long length behaviour in terms of critical phenomena. Hence the
basic properties of such polymers are argued to display scaling
behaviour. Much work has been subsequently done to verify specific
scaling predictions in both two and three dimensions (for examples
from the past 10 years see
\cite{prellberg1994a-:a,grassberger1995a-a,grassberger1995b-a,prellberg1995b-:a,grassberger1997a-a,grassberger2002a-a,hsu2002a-a}).
More generally, scaling usually imposes certain relationships between
critical exponents (a review of the more general scaling framework can
be found in \cite{brak1993a-:a} -- see also
\cite{janse2000a-a}) and it is these relationships that this Comment
addresses.

Of particular interest to this Comment is the scaling of quantities on
approaching the $\theta$-point from the collapsed phase. The collapsed
phase itself has received attention relatively recently
\cite{owczarek1993e-:a,owczarek1993d-:a,grassberger2002a-a,hsu2002a-a}.
Much less is known here, partially because the long length behaviour
is no longer a critical phenomenon. This is in contrast to the swollen
phase and the transition point, which are both critical.  To make
clear the answers to the questions raised in GH
and HG it is first
timely to restate, in compact fashion, the conjectured crossover
scaling theory for a single polymer between good and poor solvents
(high and low temperatures respectively), and then demonstrate how
questions such as those raised can be answered in general.


As temperature gets decreased, an isolated polymer in solution undergoes a 
phase transition from a swollen coil to a collapsed globule via a critical 
$\theta$-state at a temperature $T_\theta$. The standard description of this 
polymer collapse transition is a tricritical point related to the $n
\rightarrow 0$ limit of the $(\phi^2)^2$--$(\phi^2)^3$ O($n$) field theory
\cite{gennes1975a-a,stephen1975a-a,duplantier1982a-a}. Scaling theory can
therefore be derived in principle from this tricriticality
\cite{lawrie1984a-a}.
The upper critical dimension for the swollen state is four while for
the $\theta$-state it is expected to be three. As confluent
logarithmic corrections complicate the discussion in three dimensions,
the crossover theory should be cleanest for dimensions strictly below
the upper critical dimension. 
Let us therefore concentrate our
discussions on two dimensions.  

Consider now, for simplicity, some
quantity $Q(T,N)$, associated with a property of the polymer, that is
a function of the length $N$ and the temperature $T$ of the
polymer. Moreover, let it be a quantity that has an algebraic
asymptotic behaviour for large $N$ at any fixed value of $T$, such as
the radius of gyration $R_g(T,N)$ for example. Such a quantity would
then be expected to possess three different behaviours: For fixed $T>
T_\theta$
\begin{equation}
\label{q-high}
Q(T,N) \sim a_{+}(T) \: N^{q_+}\;,
\end{equation} for $T < T_\theta$
\begin{equation}
\label{q-low}
Q(T,N) \sim a_{-}(T) \: N^{q_-}\;,
\end{equation}
while for $T =T_\theta$
\begin{equation}
\label{q-theta}
Q(T,N) \sim a_{\theta} \: N^{q_\theta}\;,
\end{equation}
each as $N \rightarrow \infty$.  The assumption of crossover scaling
\cite{lawrie1984a-a,griffiths1973a-a,brak1995a-:a} applied to this
system \cite{brak1993a-:a} implies that there exists a crossover
exponent $\phi$ such that for each fixed value of $x= t N^\phi$ where
$t = \frac{T - T_\theta}{T_\theta}$
\begin{equation}
Q(T,N) \sim N^{q_\theta} G( t N^\phi) \text{ as  } N \rightarrow \infty\;.
\label{q-crossover}
\end{equation}
(Note that \cite{brak1993a-:a} contains numerous typesetting errors in
some formulae, such as eqn.\ (26), that makes readability less than
optimal --- see Ch.\ 2 in \cite{janse2000a-a} for a nice summary.)
Moreover, and importantly, it is assumed
\cite{brak1995a-:a} that this asymptotic form provides all the
dominant asymptotics for small $t$ so that
\begin{equation}
G(x)\sim\left\{
\begin{array}{ll}
b_+ \: x^{(q_+ - q_\theta)/\phi} &\quad x \rightarrow \infty \;,\\
b_- \:(-x)^{(q_- - q_\theta)/\phi}  &\quad x \rightarrow - \infty\;.
\end{array}\right.
\end{equation}
This means that the high and low temperature forms, eqns.\
(\ref{q-high}) and (\ref{q-low}) respectively, are recovered in the
appropriate limits. Consequently, it further implies that
\begin{equation}
a_{+}(T) \sim b_+ \: t^{(q_+ - q_\theta)/\phi} \text{ as } t
\rightarrow 0^+
\end{equation}
and 
\begin{equation}
\label{a-low}
a_{-}(T) \sim b_+ \: (-t)^{(q_- - q_\theta)/\phi}\text{ as } t \rightarrow 0^-\;.
\end{equation}
Since these exponents often have separate definitions a crossover
theory provides relationships between exponents defined by asymptotics
in $N$ at $T=T_\theta$ and asymptotics in $t$ on approaching
$T_\theta$. We note that suitable adjustments can sometimes be made to
this scenario if some or all of the asymptotic behaviours are not
algebraic, for example in the scaling of the partition function where
exponential as well as algebraic factors arise \cite{brak1993a-:a}.

	Armed with the general principle of crossover scaling
described above it is a simple matter to deduce answers to the
questions posed in GH. Firstly, the density
$\hat{\rho}$ inside the collapsed polymer is considered on approaching
$T_\theta$ from below. Since the density is defined as
\begin{equation}
\hat{\rho}(T) = \lim_{N \rightarrow \infty} \rho(T,N) = \lim_{N \rightarrow
\infty} \frac{N}{R_g(T,N)^d}\;,
\end{equation}
where $R_g(T,N)$ is the radius of gyration (which is a quantity
obeying the rules above for a general quantity $Q(T,N)$), we can then
relate how $\hat{\rho}(T)$ behaves on approaching $T_\theta$. Because
the polymer is expanded at high temperatures, $\hat{\rho}(T)$ is zero
for $T \geq T_\theta$. For $T < T_\theta$, $R_g(T,N) \sim r_{-}(T) \: N^{1/d}$ so that
$\rho(T,N) \sim {r_{-}(T)}^{-d}$. In fact,
$\hat{\rho}(T)$ acts like an order parameter for the transition just as
the density is an order parameter in a liquid-gas transition. Let us
define $\beta$ as
\begin{equation}
\hat{\rho}(T) \sim  (-t)^\beta \text{ as } t \rightarrow 0^-
\end{equation}
and $\nu_\theta$ via $R_g(T_\theta,N) \sim r_{\theta} \:
N^{\nu_\theta}$.  Hence, using the crossover theory above, we have
from the analogous formula to (\ref{a-low})
\begin{equation}
r_{-}\sim b_- \: (-t)^{(1/d - \nu_\theta)/\phi} \text{ as } t \rightarrow 0^-
\end{equation}
and therefore
\begin{equation}
\label{beta-value}
\beta = \frac{(d\nu_\theta -1)}{\phi}\;.
\end{equation}
In two dimensions (\ref{beta-value})  implies that $\beta =1/3$, since
$\nu_\theta = 4/7$ and $\phi =3/7$ are expected
\cite{duplantier1987a-a}. This answers the question raised in section
II of GH. It also is well supported by the
numerical evidence provided in HG, where it is
estimated that $\beta_{2d} = 0.32$.

Also of interest in GH and HG
is the scaling of the (reduced) surface free energy
$s(T)$ in the collapsed phase. (Again $s(T) =0$ for $T \geq
T_\theta$.) $s(T)$ can be defined (up to a factor dependent on the
average shape of the surface) via the scaling of the polymer partition
function in low temperatures,
\cite{owczarek1993e-:a,owczarek1993d-:a}
\begin{equation}
\label{low-part-scale}
Z(T,N) \sim e^{f(T) N + s(T) N^{(d-1)/d} + A_-(T)}  N^{\gamma_- - 1}\;,
\end{equation}
where $f(T)$ is the bulk (reduced) free energy per monomer, $A_-$ is a
non-constant function of $T$, and $\gamma_-$ may also be non-constant
(to simplify the discussion below we will assume it is constant).  Let
us define $\chi$ via
\begin{equation}
s(T) \sim c_- \: (-t)^\chi \text{ as } t \rightarrow 0^-\;.
\end{equation}
Matching the crossover scaling form for the partition function,
\begin{equation}
\label{cross-part-scale}
Z(T,N) \sim e^{f(T_\theta) N} N^{\gamma_\theta - 1} H(t N^\phi)\;,
\end{equation}
with (\ref{low-part-scale}) implies that
\begin{equation}
H(x) \sim (-x)^{-(\gamma_\theta - \gamma_-)/\phi} e^{c_1 \:
(-x)^{\frac{1}{\phi}} + c_2 \:(-x)^\frac{(d-1)}{d \phi}}
\end{equation}
as $x \rightarrow -\infty$ with $c_1$ and $c_2$ constants.
This immediately gives
\begin{equation}
\label{chi}
\chi = \frac{(d-1)}{d \phi}\;.
\end{equation}
So for two dimensions $\chi = 7/6$ holds. At this point in the
discussion it is important to note that $\chi$ will be difficult to
estimate as it requires an accurate estimate of $s(T)$, which is part
of a sub-dominant factor in the scaling of the partition function (or
free energy), just in the region of temperature affected by strong
crossover effects from the change in the $\gamma$ exponent. We remark
that the same result (\ref{chi}) can be found from the scaling Ansatz
for a suitably defined finite sized surface free energy $s(T,N)$.
Assuming $s(T,N) \sim N^{-(d-1)/d} K(t N^\phi)$ for $tN^\phi$ fixed
and then matching the fixed $T < T_\theta$ behaviour of $s(T,N)$,
namely that $\lim_{N\rightarrow\infty}s(T,N) = s(T) \neq 0$, implies
$K(x) \sim (-x)^{(d-1)/d \phi}$ as $x \rightarrow -\infty$ from which
the result again follows.

We have shown that the questions raised in recent stimulating work by
Grassberger and Hsu \cite{grassberger2002a-a,hsu2002a-a} on the
scaling of collapsing and collapsed polymers in low dimensions can be
answered by the application of crossover scaling theory. Exponents
defined by the singularity in temperature at the $\theta$-point can be
related to those defined by fixed temperature scaling in the length
and to the crossover exponent $\phi$. In particular we conjecture a
value of $1/3$ for the temperature singularity of the globular density
at the $\theta$-point in two dimensions. We also explain where the
value of $7/6$ in two dimensions arises for the temperature
singularity of the surface free energy at the $\theta$-point.

\section*{Acknowledgements} 
Financial support from the Australian Research Council is gratefully
acknowledged by ALO. ALO also thanks the Institut f\"ur Theoretische
Physik at the Technische Universit\"at Clausthal. We thank R.\ Brak
for carefully reading the manuscript and making several useful
suggestions.

\end{document}